\documentclass[12pt,onecolumn]{IEEEtran}
\usepackage{pifont}
\usepackage{bm}
\usepackage{amsfonts}
\usepackage{amssymb}
\usepackage{flushend }
\usepackage{graphicx}
\usepackage{dsfont}
\usepackage{subfigure}
\usepackage{subeqnarray}

\ifCLASSINFOpdf
\else
\fi
\usepackage[cmex10]{amsmath}
\usepackage{algorithm}

\usepackage{algorithmic}
\onecolumn 
\begin{document}
%
\title{Asymptotic Analysis of Random Lattices in High Dimensions}

\author{Rongrong~Qian and Yuan Qi\\ Beijing Univ. of Posts and
Telecommunications
(BUPT), \\
Beijing, China \\ e-mail: RongrongQian@bupt.edu.cn
\thanks{The authors are with Wireless Signal Processing and Network Lab
(Key Lab. of Universal Wireless Communication, Ministry of
Education), Beijing Univ. of Posts and Telecommunications (BUPT),
Beijing, China.} \thanks{This work is sponsored by the National Natural Science Foundation of China (grant no. 61501043).}}

%



\maketitle

\begin{abstract}
This paper presents the asymptotic analysis of random lattices in high dimensions to clarify the distance properties of the considered lattices. These properties not only indicate the asymptotic value for the distance between any pair of lattice points in high-dimension random lattices, but also describe the convergence behavior of how the asymptotic value approaches the exact distance. The asymptotic analysis further prompts new insights into the asymptotic behavior of sphere-decoding complexity and the pairwise error probability (PEP) with maximum-likelihood (ML) detector for a large number of antennas.



%

\end{abstract}


%
\IEEEpeerreviewmaketitle

\section{Introduction}
%
%
%
%


\IEEEPARstart{T}here has been a great deal of research over the past several decades on the lattice theory \cite{Pohst: Improved}-\cite{Wubben:lattices}. The studies span multiple disciplines and include mathematics \cite{Pohst: Improved}-\cite{Schnorr Euchner:subset}, information theory \cite{Agrell:Closest}-\cite{Korkine}, communication \cite{Zhao:random lattices}\cite{Damen:space-time codes}, and signal processing \cite{Hassibi:sphere decoding}\cite{Wubben:lattices}. In lattice theory, previous efforts have been mostly on or related to the \emph{closest point search} (CPS) problem and its low-complexity algorithms. Pioneering works \cite{Pohst: Improved}-\cite{Schnorr Euchner:subset} have laid a firm foundation for the Fincke-Pohst algorithm (a.k.a. sphere-decoding algorithm) and the Schnorr-Euchner algorithm, which became the mainstream of the CPS algorithms. The contributions of these works are of decisive importance, but the subject has not ended with them.


Recent years have witnessed a growing interest in the CPS problem and its algorithms, while this interest was intensified by the connection between CPS and maximum-likelihood (ML) detection in multiple-input multiple-output (MIMO) channels \cite{Mow:Maximum}\cite{Damen:closed lattice}\cite{Gamal:diversity-multiplexing tradeoff}\cite{Jalden:exponent}\cite{Damen:space-time codes}. In principle, one can represent the MIMO environment by a lattice sphere packing, and applying the universal lattice decoder in a MIMO system \cite{Damen:space-time codes}. Nowadays, the emerging large-MIMO systems which rely on very large antenna arrays have become a hot topic of communications, because as the demands on data rate and throughput increase dramatically, the number of antennas needs to be scaled up to tens or hundreds to fulfill performance requirements \cite{Ma:Element}\cite{Hanzo:Fifty}.

Many studies are performed for the algorithm design of large-MIMO detection (see \cite{Rajan:large}\cite{Ma:Element} and references therein) and the corresponding large-system performance and complexity analysis \cite{Liang:Asymptotic}. Jalden and Ottersten \cite{Jalden:digital}, Liang \emph{et al.} \cite{Liang:Asymptotic}, Evans and Tse \cite{Tse:Large system performance}, Biglieri \emph{et al.} \cite{Biglieri:a large number of antennas}, Loyka and Levin \cite{Loyka:large MIMO systems} have gained deep insights into the large-MIMO detection and associated system performance very early, even before the benefit of large-MIMO systems was widely recognized, which shows their impressive foresight.

\emph{Motivations of this paper}: First, the overwhelming majority of existing works concern large-MIMO systems and detection\cite{Jalden:digital}\cite{Liang:Asymptotic}\cite{Ma:Element}\cite{Hanzo:Fifty}\cite{Tse:Large system performance}-\cite{Loyka:large MIMO systems}, but little attention was paid to the lattices in high dimensions, even though the large-MIMO systems are closely related to the lattices in high dimensions. The points of interest of lattice theory are, after all, not entirely the same as those of MIMO system. Hence, for the sake of completeness of the lattice theory, it is important to investigate the lattices in high dimensions, which is not to simply duplicate works already done in the large-MIMO systems, but to obtain new theoretical results from a unique perspective of lattice theory.

Second, the studies in \cite{Jalden:digital}\cite{Biglieri:a large number of antennas}\cite{Loyka:large MIMO systems} resort to asymptotic analysis to approximate the exact performance of large-MIMO system and the complexity of large-MIMO detection with the asymptotic performance and complexity, respectively, because the asymptotic performance and complexity can usually be expressed in closed-form. However, these works seldom figure out what the convergence behaviors will be as the asymptotic values approach the exact ones. Thus, of particular importance now is to clarify these convergence behaviors.

Third, by far, the most widely known limit of the sphere-decoding algorithm is the exponential complexity in large systems \cite{Jalden:digital}\cite{differential detection}. Jalden and Ottersten have shown that the complexity of the sphere-decoding algorithm is exponential in the dimension $m$ of the transmitted symbol vector when applied to MIMO detection. This is sketchily due to the fact that the sphere radius has to grow linearly with $m$ to ensure that the transmitted signal is found inside the sphere with non-vanishing probability also for large $m$ \cite{differential detection}. It is worth attempting to find a more intuitive explanation for the exponential complexity of the sphere-decoding algorithm, and then seek new methods of reducing the complexity of the algorithm to be subexponential or even polynomial without sacrificing other performances.


\emph{Contributions}: First, we present the asymptotic analysis of random lattices in high dimensions to clarify the distance properties of the considered lattices. To analyze the distance properties for lattices is not a trivial task, especially for random lattices, and it seems unlikely that closed-form expressions for any but trivial systems will exist. In this paper, we derive the Chernoff bound related expressions of the distance properties (\textbf{Theorem 1}, \textbf{Corollaries 1, 2}). These properties, on the one hand, indicate an asymptotic value of the distance between any pair of lattice points in high-dimension random lattices (\textbf{Corollary 1}), and on the other hand, describe the convergence behaviors of how the asymptotic values approaches the exact distances (\textbf{Theorem 1} and \textbf{Corollary 2}).

Second, the asymptotic analysis prompts new insights into the asymptotic behavior of the sphere-decoding complexity and the pairwise error probability (PEP) with ML detector for a large number of antennas. Firstly, we derive a new lower bound of the expected sphere-decoding complexity (\textbf{Theorem 2}), which applies to random (finite) lattices as well as Rayleigh-fading MIMO systems with traditional constellation schemes (e.g., QAM, PAM, and PSK). This lower bound makes us recognize that the partial cause of the exponential complexity of sphere-decoding algorithm is the codebook used to generate the lattices (\textbf{Remark 1}). Secondly, we point out the theoretical existence and a train of thought of designing the proper codebook that might be able to realize the sphere-decoding algorithm in high-dimension random lattices with subexponential or even polynomial complexity without giving rise to the decrease of other performances (\textbf{Remark 3}). Finally, we take a closer look at the convergence rate of pairwise error probability (PEP) with ML detector for a large number of antennas (\textbf{Corollary 4}).

\emph{Notations}: Matrices are set in boldface capital letters, and vectors in boldface lowercase letters. We write $a_{ij}$ for the entry in the $i$th row and $j$th column of the matrix $\bm{A}$, and $b_{i}$ for the $i$th entry of the vector $\bm{b}$. The superscripts $T$ and $\dag$ stand for the transpose and conjugate transpose, respectively. The Frobenius norm is denoted by $\| \bm{A} \|_{F} = \sqrt{\textmd{Tr}(\bm{A}^{\dag} \bm{A})} = \sqrt{\textmd{Tr}(\bm{A} \bm{A}^{\dag})}$, where $\textmd{Tr}(\cdot)$ is the trace of a square matrix. For an $n \times m$ matrix $\bm{A}$, $\bm{a}_{i}, i=1,2,\cdots,m$, denotes the $i$th column of $\bm{A}$, and the vectorization operator $\textmd{Vec}(\bm{A}) = [\bm{a}_{1}^{T} \,\,\, \bm{a}_{2}^{T}   \cdots  \bm{a}_{m}^{T}]^{T}$. $E[\cdot]$ denotes the expectation operator. We write $\overset{d}{=}$ for equality in distribution and $\overset{P}{\longrightarrow}$ for convergence \emph{in probability}. The zero-mean complex Gaussian distribution with variance $\sigma^{2}$ is
denoted by $\mathcal{CN}(0,\sigma^{2})$. $\mathbb{CZ}$ stands for the set of Gaussian integers, that is, $\mathbb{CZ} = \mathbb{Z} + \sqrt{-1} \cdot \mathbb{Z}$. The real and imaginary parts of a complex $x \in \mathbb{C}$ are denoted by $\textmd{Re}[x] \in \mathbb{R}$ and $\textmd{Im}[x] \in \mathbb{R}$, respectively. For an event $B$, let $B^{c}$ denote its complement. Given functions $f$ and $g$ of a natural number variable $n$, define the binary relation $f \asymp g$ if and only if $\underset{n \rightarrow +\infty}{\lim} \frac{\log f}{\log g} =1$, similarly, $f \overset{\smile}{\geq} g$ (or $f \overset{\smile}{\leq} g$) if and only if $\underset{n \rightarrow +\infty}{\lim} \frac{\log f}{\log g} \geq 1$ (or $\underset{n \rightarrow +\infty}{\lim} \frac{\log f}{\log g} \leq 1$). Let $\max \{a,b\}$ and $\min \{a,b\}$ denote the maximum and minimum of $a$ and $b$, respectively.

\section{Random Lattices}

An $m$-dimensional \emph{lattice} in the unitary space $\mathbb{C}^{n}$ is generated as the integer linear combination of the set of linearly independent vectors
\begin{eqnarray}
\Lambda \triangleq  \left\{  \bm{y} = \sum_{i=1}^{m} x_{i} \bm{g}_{i} \left| x_{i} \in \mathbb{CZ} \right. \right \} ,
\end{eqnarray}
where $\bm{g}_{i} \in \mathbb{C}^{n}$, and $\bm{G}=[\bm{g}_{1} \,\,\, \bm{g}_{2}   \cdots   \bm{g}_{m}]$ represents a \emph{basis} of the lattice ($\bm{G}$ is also called the \emph{generator matrix}). In the matrix form, \begin{eqnarray} \label{equ_latt_def} \Lambda = \left\{ \bm{y} = \bm{G} \bm{x} \left| \bm{x} \in \mathbb{CZ}^{m} \right. \right \}.\end{eqnarray}
It is assumed that $\kappa = n / m  \geq 1$ is a constant without loss of generality.

The random lattices to be analyzed in this paper are generated by the generator matrix $\bm{G}$ with i.i.d. zero-mean complex Gaussian $\mathcal{CN}(0,1/n)$ entries, which in communications \cite{Zhao:random lattices}\cite{Damen:space-time codes}\cite{Jalden:digital} can be used to model the received signal vectors (without being corrupted by noise) in Rayleigh-fading multiple-input multiple-output (MIMO) systems by letting $\bm{y}$, $\bm{G}$, and $\bm{x}$ be the received signal vector, channel matrix, and transmitted signal vector, respectively. Therefore, although the traditional lattice formulation is mostly constructed in Euclidean space \cite{Agrell:Closest}, we will investigate the lattices in unitary space by referring to \cite{Seethaler:distribution}\cite{Zhao:random lattices}. In fact, the results obtained by this study shall also be established for random lattices in Euclidean space. Here, note that, a Euclidean space is a finite-dimensional, real linear space with a symmetric positive-definite inner product, and a unitary space is a complex linear space with a Hermitian positive-definite inner product. Both spaces with the $\ell_{2}$ norm $\| \bm{x} \|_{2} = \left(\sum_{i=1}^{n} \left| x_{i} \right|^{2}\right)^{1/2}, $ are normed vector spaces in which the $\ell_{2}$ norm induces a metric (a notion of \emph{distance}). This metric is defined in the natural way: The distance between two vectors $\bm{p}$ and $\bm{q}$ is given by $\|\bm{p} - \bm{q}\|_{2}$.

The CPS problem refers to finding, for given lattice $\Lambda$ with a known generator $\bm{G}$ and a given input point $\bm{\widehat{y}} \in \Lambda$, a vector $\bm{x} \in \mathbb{CZ}^{m}$ such that the squared distance metric $\| \bm{\widehat{y}} - \bm{G}\bm{x} \|_{2}^{2}$ is minimized, that is,
\begin{eqnarray} \label{equ_CPS_1}
\bm{\widehat{x}}_{CPS} = \min_{\bm{x} \in \mathbb{CZ}^{m}} \left \| \bm{\widehat{y}} - \bm{G}\bm{x} \right\|_{2}^{2} =  \min_{ \bm{x} \in \mathbb{CZ}^{m}} \left \|\bm{G} \bm{\widehat{x}} + \bm{w} - \bm{G}\bm{x} \right\|_{2}^{2},
\end{eqnarray}
where $\bm{\widehat{y}} = \bm{G} \bm{\widehat{x}} + \bm{w}$ denotes a lattice point corrupted by additive Gaussian noise $\bm{w} \in \mathbb{C}^{n}$ with i.i.d. entries $w_{i} \sim \mathcal{CN}(0,N_{0})$. The solution of the CPS problem shall be greatly affected by the properties of distances between all pairs of lattice points. However, to analyze the distance properties for lattices is not a trivial task, especially for random lattices, and it seems unlikely that closed-form expressions for any but trivial systems will exist.




\section{Main Results}

We present the distance properties of random lattices in high dimension (i.e., large $m$ and $n$) via asymptotic analysis of $\|\bm{\widehat{y}} - \bm{G} \bm{x}\|_{2}^{2}$, where $\bm{\widehat{y}} = \bm{G} \bm{\widehat{x}}  + \bm{w}$.

As a column vector, $\bm{\widehat{y}} - \bm{G} \bm{x} = \bm{G} \bm{\widehat{x}}  + \bm{w} - \bm{G} \bm{x}$ satisfies
\begin{eqnarray} \label{equ_1}
\left \| \bm{G} \bm{\widehat{x}}  + \bm{w} - \bm{G} \bm{x}\right \|_{2}^{2} = \left \|   \bm{G} ( \bm{\widehat{x}} - \bm{x})  + \bm{w} \right \|_{F}^{2}.
\end{eqnarray}

By applying the singular value decomposition (SVD), \begin{eqnarray} \label{svd_1}\bm{\widehat{x}} - \bm{x} = \bm{U}_{\Delta \bm{x}} \bm{\Sigma}_{\Delta \bm{x}} \bm{V}_{\Delta \bm{x}}^{\dag},\end{eqnarray} where $\bm{U}_{\Delta \bm{x}}$ and $\bm{V}_{\Delta \bm{x}}$ are unitary matrices. Since $\bm{\widehat{x}} - \bm{x}$ can be regarded as a rank $1$ matrix, $\bm{\Sigma}_{\Delta \bm{x}}$ is a diagonal matrix with only one non-zero singular value, such that we can define
\begin{eqnarray} \label{svd_2}
\bm{\Sigma}_{\Delta \bm{x}} \triangleq [\sigma_{\Delta \bm{x}} \,\,\, 0\,\ \cdots \,\, 0]^T.
\end{eqnarray}
This further implies \begin{eqnarray} \left \| \bm{\widehat{x}} - \bm{x} \right \|_{2}^{2} = \left \| \bm{\widehat{x}} - \bm{x} \right \|_{F}^{2} = \bm{\Sigma}_{\Delta \bm{x}}^{\dag} \bm{\Sigma}_{\Delta \bm{x}} = |\sigma_{\Delta \bm{x}}|^{2}.\nonumber \end{eqnarray}


Combining (\ref{equ_1}), (\ref{svd_1}), and (\ref{svd_2}) yields
\begin{eqnarray} \label{pre_thm1}
&&  \left \|    \bm{G} (\bm{\widehat{x}} - \bm{x})  + \bm{w} \right \|_{F}^{2}  =    \left \| \left(   \bm{G} \bm{U}_{\Delta \bm{x}} \bm{\Sigma}_{\Delta \bm{x}}  + \bm{w} \bm{V}_{\Delta \bm{x}} \right) \bm{V}_{\Delta \bm{x}}^{\dag} \right \|_{F}^{2}
 \nonumber \\ && \quad =    \left \|   \bm{G} \bm{U}_{\Delta \bm{x}} \bm{\Sigma}_{\Delta \bm{x}}  + \bm{w} \bm{V}_{\Delta \bm{x}}   \right \|_{F}^{2} = \left \|   \widetilde{\bm{G}} \bm{\Sigma}_{\Delta \bm{x}}  + \widetilde{\bm{w}}   \right \|_{F}^{2}, \nonumber
\end{eqnarray}
where $\widetilde{\bm{G}} \triangleq \bm{G} \bm{U}_{\Delta \bm{x}}$ and $\widetilde{\bm{w}} \triangleq \bm{w} \bm{V}_{\Delta \bm{x}}$ are defined for notational simplicity. Then, we can get

\begin{eqnarray} \label{pre2_thm1}
 \left \|   \widetilde{\bm{G}} \bm{\Sigma}_{\Delta \bm{x}}  + \widetilde{\bm{w}}   \right \|_{F}^{2}  &=&     \left \| \widetilde{\bm{G}} \bm{\Sigma}_{\Delta \bm{x}} \right \|_{F}^{2} + \| \widetilde{\bm{w}} \|_{F}^{2} \nonumber \\
&&+ 2   \textmd{Tr} \left( \textmd{Re} \left[ \widetilde{\bm{w}}^{\dag} \widetilde{\bm{G}} \bm{\Sigma}_{\Delta \bm{x}}\right] \right).
\end{eqnarray}
From Lemma A.1, it follows that $\widetilde{\bm{G}} \overset{d}{=} \bm{G}$ and $\widetilde{\bm{w}} \overset{d}{=} \bm{w}$. Thus, $\| \widetilde{\bm{w} }\|_{F}^{2}  = \textmd{Tr}(\widetilde{\bm{w}}^{\dag} \widetilde{\bm{w}})$ is chi-squared distributed with $2n$ degrees of freedom.

The main results of this study are the following theorem and Corollaries.

\emph{Theorem 1.} The distance between $\bm{G} \bm{\widehat{x}}  + \bm{w}  \,\, (= \bm{\widehat{y}})$ and $\bm{G} \bm{x}$ satisfies
\begin{eqnarray}
 P \left(   \frac{\left \|  \bm{G} \bm{\widehat{x}} + \bm{w} - \bm{G} \bm{x} \right \|_{2}^{2}}{\left \|  \bm{\widehat{x}} - \bm{x} \right \|_{2}^{2} + n N_{0}}   \geq \theta \right)  \leq \left(\theta e^{1-\theta}\right)^{n}, \quad \nonumber
\end{eqnarray}
for the cases with $\theta > 1$; or,
\begin{eqnarray}
P \left(   \frac{\left \|  \bm{G} \bm{\widehat{x}} + \bm{w} - \bm{G} \bm{x} \right \|_{2}^{2}}{\left \|  \bm{\widehat{x}} - \bm{x} \right \|_{2}^{2} + n N_{0}}  \leq \theta \right) \leq \left(\theta e^{1-\theta}\right)^{n}, \nonumber
\end{eqnarray}
for the cases with $0 < \theta < 1$.

\begin{figure}
\normalsize
\centering
  \includegraphics[width=3.7in]{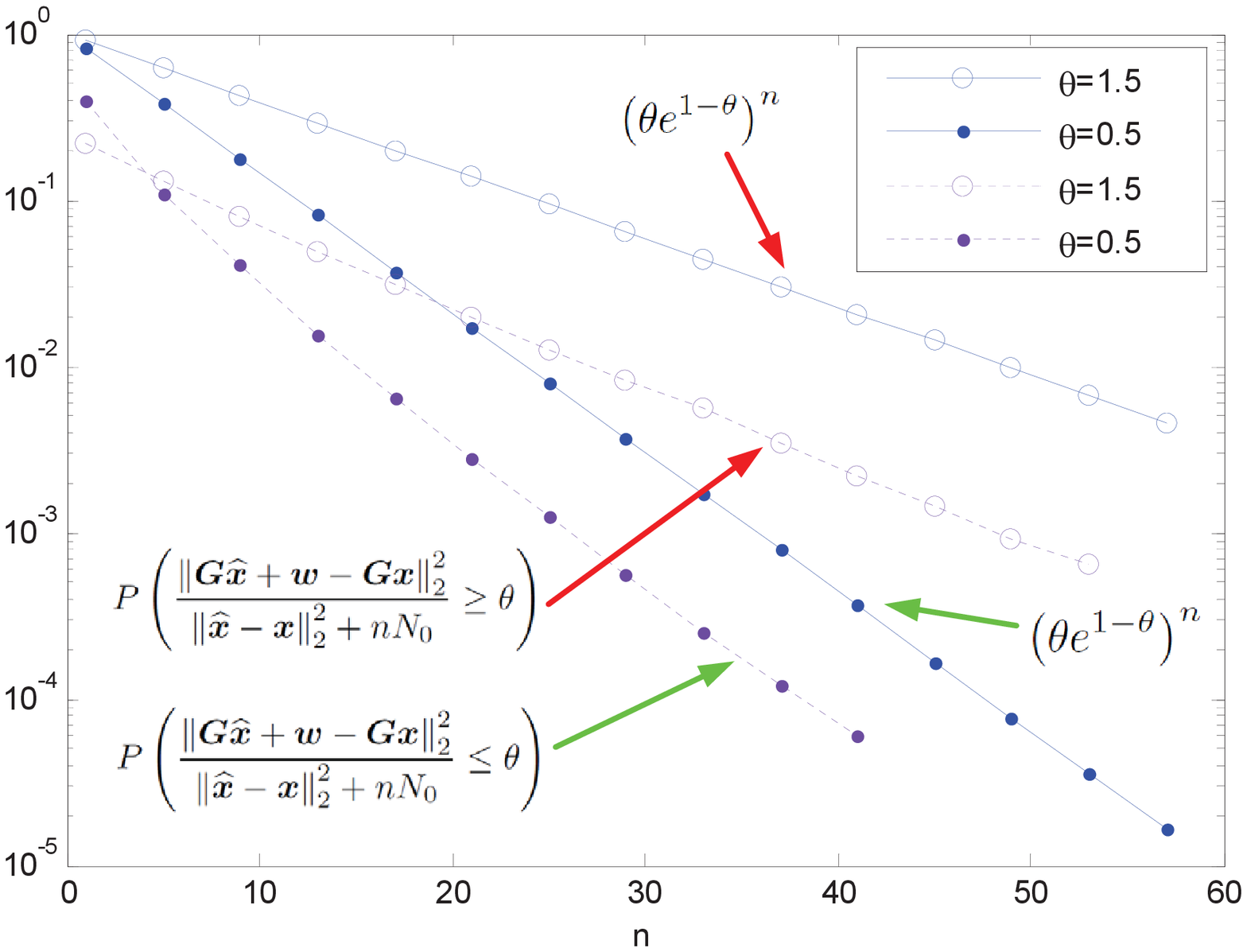}
  \caption{$P \left(   \frac{\left \|  \bm{G} \bm{\widehat{x}} + \bm{w} - \bm{G} \bm{x} \right \|_{2}^{2}}{\left \|  \bm{\widehat{x}} - \bm{x} \right \|_{2}^{2} + n N_{0}}   \geq \theta \right)$ vs its bound $\left(\theta e^{1-\theta}\right)^{n}$ for $\theta=1.5$, and $P \left(   \frac{\left \|  \bm{G} \bm{\widehat{x}} + \bm{w} - \bm{G} \bm{x} \right \|_{2}^{2}}{\left \|  \bm{\widehat{x}} - \bm{x} \right \|_{2}^{2} + n N_{0}}   \leq \theta \right)$ vs its bound $\left(\theta e^{1-\theta}\right)^{n}$ for $\theta=0.5$.} \label{Visio-cmp}
\end{figure}

\emph{Proof:} See Appendix. In addition, Fig. \ref{Visio-theta} plots the numerical results of $ \left(\theta e^{1-\theta} \right)^n$ with different $n$, and Fig. \ref{Visio-cmp} illustrates the relation between $P \left(   \frac{\left \|  \bm{G} \bm{\widehat{x}} + \bm{w} - \bm{G} \bm{x} \right \|_{2}^{2}}{\left \|  \bm{\widehat{x}} - \bm{x} \right \|_{2}^{2} + n N_{0}}   \geq \theta \right)$ and its upper bound $\left(\theta e^{1-\theta}\right)^{n}$, as well as that between $P \left(   \frac{\left \|  \bm{G} \bm{\widehat{x}} + \bm{w} - \bm{G} \bm{x} \right \|_{2}^{2}}{\left \|  \bm{\widehat{x}} - \bm{x} \right \|_{2}^{2} + n N_{0}}   \leq \theta \right)$ and its upper bound $\left(\theta e^{1-\theta}\right)^{n}$. It is observed that for both $\theta = 1.5$ and $\theta = 0.5$ the upper bound $\left(\theta e^{1-\theta}\right)^{n}$ can precisely capture the decreasing rate of $P \left(   \frac{\left \|  \bm{G} \bm{\widehat{x}} + \bm{w} - \bm{G} \bm{x} \right \|_{2}^{2}}{\left \|  \bm{\widehat{x}} - \bm{x} \right \|_{2}^{2} + n N_{0}}   \geq \theta \right)$ and $P \left(   \frac{\left \|  \bm{G} \bm{\widehat{x}} + \bm{w} - \bm{G} \bm{x} \right \|_{2}^{2}}{\left \|  \bm{\widehat{x}} - \bm{x} \right \|_{2}^{2} + n N_{0}}   \leq \theta \right)$ accompanied with the increase of $n$.

\begin{figure}
\normalsize
\centering
  \includegraphics[width=3.7in]{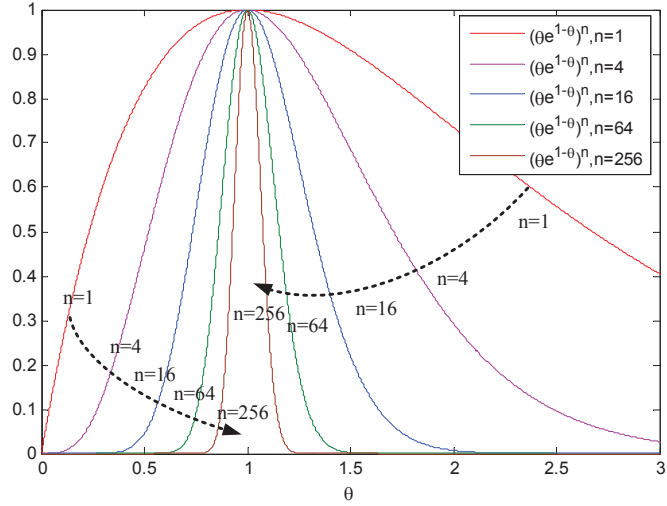}
  \caption{Numerical results of $\left( \theta e^{1-\theta} \right)^n$ with $n=1,4,16,64,256$.} \label{Visio-theta}
\end{figure}

The analytical framework employed by the proof of Theorem 1 consists of two key points: i) making use of the unitarily invariant property of $\bm{G}^{\dag}\bm{G}$ and ii) deriving the Chernoff bound of the tail probability of $ \frac{\left \| \bm{G} \bm{\widehat{x}} + \bm{w} - \bm{G} \bm{x}\right \|_{2}^{2}}{\left \|  \bm{\widehat{x}} - \bm{x} \right \|_{2}^{2} + n N_{0}}$. However, obtaining a closed-form expression of Chernoff bound is quite challenging; \cite{Proakis_Digital_1995} offers a standard approach to derive such a bound. Hence, we claim here that the aforementioned analytical framework is not all-purpose but might be a promising approach to analyze the convergence behavior of random lattices with unitarily invariant $\bm{G}^{\dag}\bm{G}$.

From Theorem 1 we can also get the below extended result.

\emph{Corollary 1.} As $n$ tends to infinity while $n/m = \kappa>1$, \begin{eqnarray} \frac{\left \| \bm{G} \bm{\widehat{x}} + \bm{w} - \bm{G} \bm{x}\right \|_{2}^{2}}{\left \|  \bm{\widehat{x}} - \bm{x} \right \|_{2}^{2} + n N_{0}}   \overset{P}{\longrightarrow} 1.  \nonumber \end{eqnarray}

\emph{Proof:} By letting $\theta=1+\varepsilon$ (or $\theta=1-\varepsilon$) where the positive $\varepsilon$ is arbitrarily close to zero, $ \frac{\left \| \bm{G} \bm{\widehat{x}} + \bm{w} - \bm{G} \bm{x}\right \|_{2}^{2}}{\left \|  \bm{\widehat{x}} - \bm{x} \right \|_{2}^{2} + n N_{0}}$ can be easily validated to converge to $1$ in probability.

At first glance, one could observe that Theorem 1 and its derivation shall establish for any positive integer $n$. Then a question arose, why this study limits the investigation to the case of large $n$. Our main concern is that only when $n$ is sufficiently large, $\left( \theta e^{1-\theta} \right)^n$ as a upper bound can determinately take (very) little value such that $\left \| \bm{G} \bm{\widehat{x}} + \bm{w} - \bm{G} \bm{x}\right \|_{2}^{2}$ can be very close to $\left \|  \bm{\widehat{x}} - \bm{x} \right \|_{2}^{2} + n N_{0}$.

Furthermore, without the presence of Gaussian noise $\bm{w}$, the inequalities in Theorem 1 can be simplified as follows.

\emph{Corollary 2.} With $\theta > 1$, it holds that
\begin{eqnarray}
 P \left(   \frac{\left \|  \bm{G} (\bm{\widehat{x}} - \bm{x}) \right \|_{2}^{2}}{\left \|  \bm{\widehat{x}} - \bm{x} \right \|_{2}^{2} }   \geq \theta \right) \leq \left(\theta e^{1-\theta}\right)^{n}. \quad \nonumber
\end{eqnarray}
With $0 < \theta < 1$,
\begin{eqnarray}
P \left(   \frac{\left \|    \bm{G} (\bm{\widehat{x}} - \bm{x})   \right \|_{2}^{2}}{\left \|  \bm{\widehat{x}} - \bm{x} \right \|_{2}^{2} }  \leq \theta \right) \leq \left(\theta e^{1-\theta}\right)^{n}. \nonumber
\end{eqnarray}

\emph{Proof:} Corollary 2 is a special case of Theorem 1 by assuming that $\bm{w} = \bm{0} \in \mathbb{C}^{n}$ and $N_{0} = 0$.

In summary, Theorem 1 and Corollary 2 are the Chernoff bound based expressions of the distance properties. Corollary 1 coarsely indicates the asymptotic value, $\left \|  \bm{\widehat{x}} - \bm{x} \right \|_{2}^{2} + n N_{0}$, for the distance between any pair of lattice points in high-dimension random lattices, $\left \|  \bm{G} \bm{\widehat{x}} + \bm{w} - \bm{G} \bm{x} \right \|_{2}^{2}$, while Theorem 1 and Corollary 2 delicately describe the convergence behaviors of how the asymptotic values, $\left \|  \bm{\widehat{x}} - \bm{x} \right \|_{2}^{2} + n N_{0}$ and $\left \|  \bm{\widehat{x}} - \bm{x} \right \|_{2}^{2} $, approach the exact distances, $\left \|  \bm{G} \bm{\widehat{x}} + \bm{w} - \bm{G} \bm{x} \right \|_{2}^{2}$ and $\left \|    \bm{G} (\bm{\widehat{x}} - \bm{x})   \right \|_{2}^{2}$, respectively. The so-called main results largely clarify several considerably abstract relations on the distance properties of random lattices in high dimension, but the benefit will in the following section be made clear by applying these results to concrete applications.

\section{Applications}

We will apply the main results obtained above to (i) the asymptotic analysis of the sphere-decoding complexity, (ii) the consideration on the codebooks of $\bm{x}$ for the sphere-decoding algorithm in high-dimension random lattices, and (iii) the convergence rate of pairwise error probability (PEP) with ML detector for a large number of antennas.

\subsection{Asymptotic Behavior of Sphere-Decoding Complexity}

Before entering into deeper discussion, we assume that $\bm{x}$ of lattice in (\ref{equ_latt_def}) comes from a finite codebook $\mathcal{C}_{\bm{x}}$ with $|\mathcal{C}_{\bm{x}}|$ codewords $\{\bm{x}(1),\cdots,\bm{x}(\mathcal{C}_{\bm{x}})\}$ and an overall power constraint on
the codebook $\mathcal{C}_{\bm{x}}$, that is, $\frac{1}{m |\mathcal{C}_{\bm{x}}|} \sum_{\bm{x} \in \mathcal{C}_{\bm{x}}} \left \| \bm{x} \right \|_{2}^{2}  =  E_{x}$ exists, such that the average power of each entry in $\bm{x}$ is $E_{x}$. In communications, the problem of (\ref{equ_CPS_1}) is known as the maximum-likelihood (ML) detection problem for which $\bm{x}$ is always assumed to be drawn from a codebook as \cite{Gamal:diversity-multiplexing tradeoff}\cite{Jalden:exponent}\cite{Damen:space-time codes} \begin{eqnarray}\label{equ_constell} \mathcal{C}_{\bm{x}} = \mathcal{C}_{\tau,L}^{m} \triangleq \{x | \textmd{Re} [x], \textmd{Im} [x] \in \mathbb{R} \cap [\tau \,\, \, \tau + L-1]\}^{m}, \end{eqnarray} where $L$ regulates the set size for each entry of $\bm{x}$. In the remainder of this section, we suppose that $\bm{x}$ uses the codebook of (\ref{equ_constell}) by default, unless explicitly stated.

The worst-case complexity for solving the CPS problem optimally for generic lattices is NP-hard, while the search of lattice points can be realized efficiently by sphere decoding \cite{Viterbo Boutros:universal}\cite{Damen:space-time codes}. The main idea of the sphere-decoding algorithm for solving the problem (\ref{equ_CPS_1}) is based on enumerating all points $\bm{x}$ such that $\bm{Gx}$ lies within a sphere of radius $\rho$ centered at $\bm{\widehat{y}}$, that is, on enumerating all $\bm{x}$ subject to the \emph{sphere constraint} (SC)\cite{Seethaler:distribution}\cite{Jalden:digital}
\begin{eqnarray} \label{equ_SD_2}
 &&   \left \| \bm{G} \bm{\widehat{x}}  + \bm{w} - \bm{G} \bm{x}\right \|_{2}^{2} \leq \rho^{2},
\end{eqnarray}
and then choosing the one that minimizes the distance metric.

To make use of the Theorem 1 and Corollary 1, define
\begin{eqnarray}
&&N_{SC}(\bm{G},\bm{\widehat{x}},\bm{w}) \triangleq  \left | \left \{\bm{x} \in \mathbb{CZ}^{m} \left| \left \| \bm{G} \bm{\widehat{x}}  + \bm{w} - \bm{G} \bm{x}\right \|_{2}^{2} \leq \rho^{2} \right. \right \} \right | , \nonumber \\
 &&N'(\bm{\widehat{x}},\theta)  \triangleq   \left| \left \{\bm{x} \in \mathbb{CZ}^{m} \left| \left \|  \bm{\widehat{x}} - \bm{x} \right \|_{2}^{2} + n N_{0} \leq \frac{\rho^2}{\theta} \right. \right \} \right|. \nonumber
\end{eqnarray}

\emph{Corollary 3.} If it holds that $E \left[N'(\bm{\widehat{x}},\theta)\right] \overset{\smile}{\geq} e^{n \psi}$ with $\theta > 1$ and $\psi>0$, then
\begin{eqnarray} \label{equ_cor_3a}
E \left[ N_{SC}(\bm{G},\bm{\widehat{x}},\bm{w}) \right] \overset{\smile}{\geq} e^{n \psi}.
\end{eqnarray}
or, if $E \left[N'(\bm{\widehat{x}},\theta)\right] \overset{\smile}{\leq} e^{n \psi}$ with $\theta < 1$ and $\psi>0$,
\begin{eqnarray} \label{equ_cor_3b}
E \left[ N_{SC}(\bm{G},\bm{\widehat{x}},\bm{w}) \right] \overset{\smile}{\leq} e^{n \psi_{max}}.
\end{eqnarray}
where we assume that $L^{m}$ is the worst-case complexity of sphere-decoding without loss of generality, and
\begin{eqnarray}
\psi_{max} = \max \left\{ \psi, \log \theta + 1 - \theta + \frac{\log L}{\kappa}  \right\}. \nonumber
\end{eqnarray}

\emph{Proof:} See Appendix.

The calculation of the metric constraint (\ref{equ_SD_2}) can also be written, after QR factorization of $\bm{G}$, as
\begin{eqnarray} \label{equ_SD_3}  \left \| \bm{R} \bm{\widehat{x}}  + \bm{w}' - \bm{R} \bm{x}\right \|_{2}^{2} \leq \rho^{2}, \nonumber \end{eqnarray} where $\bm{R}$ is an $m \times m$ upper triangular matrix with positive real-valued elements on its main diagonal and $\bm{w}' = \bm{Q}^{H} \bm{w}$, by assuming $\bm{G} = \bm{QR}$ such that $\bm{Q}$ is unitary of
dimension $n \times m$. The metric $ \left \| \bm{R} \bm{\widehat{x}}  + \bm{w}' - \bm{R} \bm{x}\right \|_{2}^{2}$ can be computed recursively by
\begin{eqnarray}
&& \left \| \bm{R}_{k} \bm{\widehat{x}}_{k}  + \bm{w}'_{k} - \bm{R}_{k} \bm{x}_{k} \right \|_{2}^{2} \nonumber \\ && \quad = \left \| \bm{R}_{k-1} \bm{\widehat{x}}_{k-1}    + \bm{w}'_{k-1} - \bm{R}_{k-1} \bm{x}_{k-1} \right \|_{2}^{2}  \nonumber \\ &&  \quad \quad + \left|w'_{m-k+1} + \sum_{i=m-k+1}^{m} r_{m-k+1,i} \left(\widehat{x}_{i} - x_{i}\right) \right|^2, \nonumber \end{eqnarray}
where
$\bm{x}_{k} = [x_{m-k+1}  \cdots  x_{m}]^{T}$, $\bm{\widehat{x}}_{k} = [\widehat{x}_{m-k+1}   \cdots  \widehat{x}_{m}]^{T}$, $ \bm{w}'_{k} = [w_{m-k+1}'  \cdots  w_{m}']^{T}$, and $\bm{R}_{k}$ refers to the $k \times k$ bottom right (upper triangular) submatrix of $\bm{R}$ associated with $\bm{\widehat{x}}_{k}-\bm{x}_{k}$.

This means that the sphere-decoding algorithm is able to identify whether the lattice points locate in the considered sphere by using the \emph{recursive} sphere constraint (RSC) \begin{eqnarray} \label{equ_SD_4}  \left \| \bm{R}_{k} \bm{\widehat{x}}_{k}  + \bm{w}'_{k} - \bm{R}_{k} \bm{x}_{k} \right \|_{2}^{2} \leq \rho^2,\end{eqnarray} starting from $k=1$ to $m$, and finally, ending with $\left \| \bm{R}_{m} \bm{\widehat{x}}_{m}  + \bm{w}'_{m} - \bm{R}_{m} \bm{x}_{m}\right \|_{2}^{2} = \left \| \bm{R} \bm{\widehat{x}}  + \bm{w}' - \bm{R} \bm{x}\right \|_{2}^{2}  \leq \rho^2$.

As it is customary in the literature (cf. \cite{Seethaler:distribution}\cite{Jalden:digital}), for given $\bm{G}$, $\bm{\widehat{x}}$, and $\bm{w}$, the sphere-decoding complexity, $C_{SD}(\bm{G},\bm{\widehat{x}},\bm{w})$, is usually defined as the number of lattice points searched by the algorithm, that is, the total number of vectors $\bm{x}_{k}, k = 1,\cdots,m$, that satisfies the RSCs in (\ref{equ_SD_4}), such that \begin{eqnarray} C_{SD}(\bm{G},\bm{\widehat{x}},\bm{w}) = \sum_{k=1}^{m} N_{k}(\bm{R}_{k},\bm{\widehat{x}}_{k},\bm{w}'_{k}),\nonumber\end{eqnarray} where
\begin{eqnarray} \label{equ_SD_ck}
&& N_{k}(\bm{R}_{k},\bm{\widehat{x}}_{k},\bm{w}'_{k}) \triangleq \nonumber \\ && \quad   \left| \left \{\bm{x}_{k} \in \mathbb{CZ}^{k} \left| \| \bm{R}_{k} \bm{\widehat{x}}_{k} + \bm{w}'_{k} - \bm{R}_{k} \bm{x}_{k} \|_{2}^{2}   \leq \rho^2 \right. \right \} \right|. \quad \quad \nonumber
\end{eqnarray}

The expected complexity of the sphere-decoding algorithm is thereafter computed by
\begin{eqnarray}
C_{SD} = E \left[C_{SD}(\bm{G},\bm{\widehat{x}},\bm{w})\right]. \nonumber
\end{eqnarray}
By definition, we directly obtain a lower bound of $C_{SD}$ as

\begin{eqnarray} \label{equ_CSD_lb}C_{SD} \geq E[ N_{SC}(\bm{G},\bm{\widehat{x}},\bm{w}) ]. \nonumber\end{eqnarray}

A general agreement in the community exists that the sphere-decoding complexity can be reduced by employing the preprocessing techniques such as lattice-reduction (LR) or layer-sorting (LS) \cite{Wubben:lattices}; however, $E[N_{SC}(\bm{G},\bm{\widehat{x}},\bm{w})]$ is only determined by the SC in (\ref{equ_SD_2}) and shall not be affected by these techniques. In essence, most preprocessing techniques aim to cut down $\sum_{k=1}^{m-1} N_{k}(\bm{R}_{k},\bm{\widehat{x}}_{k},\bm{w}'_{k})$ for $C_{SD}(\bm{G},\bm{\widehat{x}},\bm{w})$, and thus, $E[ N_{SC}(\bm{G},\bm{\widehat{x}},\bm{w}) ]$ represents a theoretical limit to which $C_{SD}$ tries to approach. Note that this study did not consider the early-termination strategies or adaptive radius-updating strategies which are heuristic methods of complexity reduction for the sphere-decoding algorithm and could impact $E[N_{SC}(\bm{G},\bm{\widehat{x}},\bm{w})]$.

\emph{Theorem 2.} If let $\rho^2 = \alpha n N_{0}$ with $\alpha > 1$, a closed-form lower bound of $C_{SD}$ is obtained as \begin{eqnarray}\label{equ_lb_our} C_{SD} \overset{\smile}{\geq} L^{n \cdot \min \left \{1/\kappa, \frac{(\alpha-1) N_{0}}{ d_{max}^2 } \right \} },\end{eqnarray}
where it is assumed that $n/m = \kappa > 1$ as in the preceding, $d_{max}^2$ denotes the maximum distance of codewords in $\mathcal{C}_{\tau,L}$ so that
 \begin{eqnarray}
d_{max}^2 = \max_{a, b \in \mathcal{C}_{\tau,L}} |a - b|^2. \nonumber
\end{eqnarray}

\emph{Proof:} See Appendix.

It shall be emphasized that Theorem 2 (with slight modifications) can be applied to Rayleigh-fading MIMO systems with traditional constellation schemes, for example, QAM, PAM, and PSK \cite{Proakis_Digital_1995}. Because by mapping the elements $x \in \mathcal{C}_{\tau,L}$ to elements $s \in \mathbb{S}$ using the transformation $s = ax+b$, the CPS problem is equivalent to a MIMO detection problem in wireless communications \cite{Gamal:diversity-multiplexing tradeoff}\cite{Jalden:exponent}\cite{Damen:space-time codes}, where $\mathbb{S}$ denotes the set of constellation used by the multiple-antenna systems.

For the sphere-decoding algorithm in digital communications, the expected complexity is proved in \cite{Jalden:digital} to be lower bounded by an exponential function of $L$, in considering the $L$-PAM constellation and Rayleigh-fading channel matrix $\bm{H}$ with a power constraint as $E[\|\bm{h}_{i}\|_{2}^{2}] \leq c^2, \forall i \in [1 \,\, m]$. If using the notations of this study, the above bound can be written as
 \begin{eqnarray} \label{equ_lb_jj}
C_{SD} \geq \frac{L^{\eta m}-1}{L-1}, \quad \eta = \frac{1}{2} \left(\frac{c^2 (L^2-1)}{3 N_{0}} + 1 \right)^{-1}.
\end{eqnarray}

We hasten to compare (\ref{equ_lb_our}) and (\ref{equ_lb_jj}) for clarifying their difference and relevance. To be specific,

\begin{itemize}
  \item The derivation of Theorem 2 offered in this paper makes less assumptions (constraints) on $\bm{x}$ and $\bm{G}$ than $\bm{s}$ and $\bm{H}$ accordingly in \cite{Jalden:digital}, thus, the formulation of (\ref{equ_lb_our}) applies to more systems than (\ref{equ_lb_jj}).
  \item In case of high SNR (e.g., $N_{0}$ is very small while $d_{max}^2$ and $L^2-1$ care kept relatively large), the ratio $\frac{N_{0}}{d_{max}^2}$ and $\frac{N_{0}}{L^2-1}$ dominate the exponents of lower bounds in (\ref{equ_lb_our}) and (\ref{equ_lb_jj}), respectively. Here, the maximum distance $d_{max}^2$ has tight connection with $L^2-1$ because $d_{max}^2$ of codewords in the $L$-PAM constellation is $(L-1)^2$ that shall be close to $L^2-1$ especially for large $L$.
        \item It might not be difficult to explain why the condition $E[\|\bm{h}_{i}\|_{2}^{2}] \leq c^2, \forall i \in [1 \,\, m]$ is used by \cite{Jalden:digital} but not needed in this study. As $n \rightarrow + \infty$, $\|\bm{h}_{i}\|_{2}^{2} \rightarrow 1$ in probability. Then, if with $c > 1$, $E[\|\bm{h}_{i}\|_{2}^{2}] \leq c^2$ establishes by default for large $n$.
\end{itemize}

\subsection{Codebook of $\bm{x}$ for Sphere-Decoding Algorithm in High-Dimension Random Lattices}

As pointed out earlier, the expected complexity of the sphere-decoding algorithm has exponential lower bound (as in Theorem 2, or in \cite[Theorem 2]{Jalden:digital}) under the given assumptions. However, it is not fair to totally ascribe the cause of the exponential complexity to the sphere-decoding algorithm. We will indicate and draw attention to a hitherto unnoticed point on the cause of exponential complexity.

  \emph{Remark 1.} It is reasonable to attribute the partial cause of the exponential complexity of sphere-decoding algorithm to the codebook of $\bm{x}$, because first of all, two facts shall not be ignored, which are given as follows:
  \begin{enumerate}
    \item Letting $\rho^2 = \alpha n N_{0}$ implies that the radius of SC would grow with $n$, while the codewords in the codebook $\mathcal{C}_{\tau,L}^{m}$ of (\ref{equ_constell}) has the fixed minimal distance whatever the value of $n$ is. That is, as $n$ increases, the sphere-decoding algorithm will search for the closet lattice point in a sphere with larger boundary; in contrast, the minimal distance of the codewords of $\bm{x}$ does not change with $n$.
  \item The main results state that as $n$ grows large, the distance metric $\left \| \bm{G} \bm{\widehat{x}}  + \bm{w} - \bm{G} \bm{x}\right \|_{2}^{2}$ tends to $\left \|  \bm{\widehat{x}} - \bm{x} \right \|_{2}^{2} + n N_{0}$ so that $\left \|  \bm{\widehat{x}} - \bm{x} \right \|_{2}^{2}$ is going to be the dominated factor of the distance between any pair of lattices (probably corrupted by noise) while the effects of $\bm{G}$ and $\bm{w}$ gradually vanish.
  \end{enumerate}
From these, one can imagine that the complexity of the sphere-decoding algorithm will rise dramatically as the dimension of lattices becomes high; during this procedure the codebook of $\bm{x}$ becomes the dominated factor of the sphere-decoding complexity.

To simplify the presentation, let us introduce a definition that describes how the minimal distance of the codes of $\bm{x}$ performs for high-dimension lattices.

\emph{Definition 1.} Given a non-decreasing function of $m$, $\gamma(m)$, the codebook of $\bm{x}$ is said to \emph{inflate} with $\gamma(m)$ as $m$ increases, if there exists one monotonically increasing function $g(\cdot) > 0$, such that, $\exists m' \in \mathbb{Z} \cap (0 \,\, +\infty)$, for $\forall m \geq m'$,
\begin{eqnarray}
d_{min}(m)\geq g(\gamma(m)), \nonumber
\end{eqnarray}
where $d_{min}(m)$ denotes the minimal distance of the codebook of $\bm{x}$. Then, we say that the codebook of $\bm{x}$ is \emph{not} inflatable, if, for $\forall m \in \mathbb{Z} \cap (0 \,\, +\infty)$ there exists a positive $\overline{c}$,

\begin{eqnarray}
d_{min}(m) \leq \overline{c}. \nonumber
\end{eqnarray}

\emph{Remark 2.} Clearly, the codebook $\mathcal{C}_{\tau,L}^{m}$ in (\ref{equ_constell}) is not inflatable, which in turn implies that the traditional constellation schemes in Rayleigh-fading MIMO systems corresponds to the codebooks (of $\bm{x}$) that are also not inflatable. More suitable codebooks of $\bm{x}$ for the sphere-decoding algorithm in high-dimension random lattices may be obtained by considering that, if the minimal distance $d_{min}(m)$ of some codebooks of $\bm{x}$ can inflate with the SC radius of the sphere-decoding algorithm, $\rho(m)$, where $\rho(m)  = \sqrt{\alpha n N_{0}} = \sqrt{\alpha \kappa m N_{0}}$, then the minimal distance $d_{min}(m)$ and $\rho(m)$ can increase with $m$ simultaneously so that the number of lattice points satisfying the SC of the sphere-decoding algorithm might grow subexponentially or even polynomially with $m$.

In what follows, the existence of a potential code scheme that inflates with $\rho(m)$ will be discussed theoretically. Let $V^{\mathbb{C}}_{1,m} \triangleq \{\Phi \in \mathbb{C}^{m \times 1} | \Phi^{\dag} \Phi = 1\}$ denote the (complex) Stiefel manifold. The canonical embedding of $V^{\mathbb{C}}_{1,m}$ into the vector space $\left( \mathbb{C}^{m \times 1}, <\cdot,\cdot>_{\mathbb{C}} \right)$ motivates the definition of the topological `chordal' metric/distance $d(\Phi,\Psi) = \| \Phi-\Psi \|_{F}$, $\Phi,\Psi \in V^{\mathbb{C}}_{1,m}$ \cite{Henkel:Sphere packing}.

\emph{Theorem (Sphere-packing bound \cite[Corollary IV.2]{Henkel:Sphere packing}).} For $m \gg 1$, there exist a codebook $\mathcal{C}_{\bm{x}}^{SP} \subset \{\bm{x} | \bm{x} \in  \sqrt{m E_{x}} V^{\mathbb{C}}_{1,m} \}$ with minimal distance $d_{min}(m)$ lower bounded by
\begin{eqnarray} \label{equ_18}
&& d_{min}(m) \geq \frac{\sqrt{m E_{x}}}{\alpha'} \left(\frac{1}{2}\right)^{\frac{m R}{D_{m,1}}} \nonumber \\ && \quad = \frac{\sqrt{ E_{x} / N_{0}}}{\alpha' \sqrt{\alpha \kappa}} \left(\frac{1}{2}\right)^{\frac{m R}{D_{m,1}}} \rho(m),
\end{eqnarray}
where $\alpha'$ is a constant coefficient that can bridge the chordal distance and geodesic distance as in \cite[Proposition II.1]{Henkel:Sphere packing}, $D_{m,1} = 2m-1$, and $R=\frac{1}{m} \log_{2}(|\mathcal{C}^{SP}_{\bm{x}}|)$.

To fairly compare $\mathcal{C}^{SP}_{\bm{x}}$ with $\mathcal{C}_{\tau,L}^{m}$, suppose that $|\mathcal{C}^{SP}_{\bm{x}}| = L^{m}$ such that $|\mathcal{C}^{SP}_{\bm{x}}| = |\mathcal{C}_{\tau,L}^{m}|$ and $R = \log_{2}L$. Then, (\ref{equ_18}) becomes
 \begin{eqnarray}
 && d_{min}(m) \geq \frac{\sqrt{ E_{x} / N_{0}}}{\alpha' \sqrt{\alpha \kappa L}} \rho(m). \nonumber
 \end{eqnarray}
 Besides, the codebook $\mathcal{C}_{\bm{x}}^{SP} \subset \{\bm{x} | \bm{x} \in  \sqrt{m E_{x}} V^{\mathbb{C}}_{1,m} \}$ are assumed to be derived by multiplying elements in $ V^{\mathbb{C}}_{1,m}$ with $\sqrt{m E_{x}}$ such that the power constraint $\frac{1}{m |\mathcal{C}^{SP}_{\bm{x}}|} \sum_{\bm{x} \in \mathcal{C}^{SP}_{\bm{x}}} \left \| \bm{x} \right \|_{2}^{2}  =  E_{x}$ is satisfied.

 \emph{Remark 3.} It is apparent that $\mathcal{C}_{\bm{x}}^{SP}$ inflates with $\rho(m)$, because $\frac{\sqrt{ E_{x} / N_{0}}}{\alpha' \sqrt{\alpha \kappa L}} \rho(m)$ is a monotonically increasing function of $\rho(m)$, which is in proportion to $\sqrt{ E_{x} / N_{0}}$ but inversely proportional to $\sqrt{L}$. If the codebook $\mathcal{C}_{\bm{x}}^{SP}$ can be explicitly constructed, solving the problem (\ref{equ_CPS_1}) of the random lattices may be possible using $\mathcal{C}_{\bm{x}}^{SP}$ by the sphere-decoding algorithm with the radius of $\rho(m)$ while merely expending subexponential complexity (or even polynomial complexity).

The analysis above gives a bird's eye view of the codebook of $\bm{x}$ for the sphere-decoding algorithm in high-dimension random lattices so far, focusing on the theoretical existence and a train of thought of designing the proper codebook which might be able to realize the sphere-decoding algorithm in high-dimension random lattices with subexponential or even polynomial complexity without giving rise to the sacrifice of other performances. However, the final confirmation of the existence of and how to exactly construct such a codebook are beyond the scope of this study and shall be a meaningful but challenging future work.

\subsection{Convergence Rate of Pairwise Error Probability (PEP) with ML Detector for A Large Number of Antennas}

Consider a multiple-antenna system with linear model $\bm{y} = \bm{H} \bm{x} + \bm{w}$, where $\bm{y}$ is the received signal, $\bm{x}$ is the transmitted signal, $\bm{w}$ is the noise vector of zero-mean complex Gaussian random variables with zero mean and independent real and imaginary parts with the same variance $N_{0}/2$, and $\bm{H}$ is the channel matrix whose entries are independent complex Gaussian random variables, circularly distributed with variance of their real and imaginary parts equal to $1/2n$.

The PEP is the basic building block for the derivation of union bounds to the error probability of a MIMO detector. In the considered system, the PEP with ML detector is given by

\begin{eqnarray}
P \left(\bm{\widehat{x}} \rightarrow \bm{x} \right) = E_{\bm{H}}\left[ Q\left( \frac{\left \|  \bm{H} (\bm{\widehat{x}} - \bm{x}) \right \|_{2}}{\sqrt{2 N_{0}}}\right) \right], \nonumber
\end{eqnarray}
where $Q(t) = \frac{1}{\sqrt{2 \pi}} \int_{t}^{+\infty} e^{-z^2/2} dz$ \cite{Proakis_Digital_1995}.

For a large number of antennas, we can have \cite{Biglieri:a large number of antennas}
\begin{eqnarray} E_{\bm{H}}\left[ Q\left( \frac{\left \|  \bm{H} (\bm{\widehat{x}} - \bm{x}) \right \|_{2}}{\sqrt{2 N_{0}}}\right) \right] \rightarrow Q\left( \frac{\left \|  \bm{\widehat{x}} - \bm{x} \right \|_{2}}{\sqrt{2 N_{0}}}\right).\nonumber \end{eqnarray}
This relation implies that the expectation of $Q\left( \frac{\left \|  \bm{H} (\bm{\widehat{x}} - \bm{x}) \right \|_{2}}{\sqrt{2 N_{0}}}\right)$ shall converge to $Q\left( \frac{\left \|  \bm{\widehat{x}} - \bm{x} \right \|_{2}}{\sqrt{2 N_{0}}}\right)$, however, it does not provide any more information on how fast the convergence will be.

From Corollary 2, we can further obtain the below result related to the convergence rate of $Q\left( \frac{\left \|  \bm{H} (\bm{\widehat{x}} - \bm{x}) \right \|_{2}}{\sqrt{2 N_{0}}}\right)$ to $Q\left( \frac{\left \|  \bm{\widehat{x}} - \bm{x} \right \|_{2}}{\sqrt{2 N_{0}}}\right)$.

\emph{Corollary 4.} With $\theta > 1$, it holds that
\begin{eqnarray}
 P \left( Q \left(  \frac{\left \|  \bm{H} (\bm{\widehat{x}} - \bm{x}) \right \|_{2} }{\sqrt{2 N_{0}} } \right)  \leq  Q \left( \frac{ \sqrt{\theta} \left \|  \bm{\widehat{x}} - \bm{x} \right \|_{2}}{\sqrt{2 N_{0}}} \right) \right) \leq \left(\theta e^{1-\theta}\right)^{n}. \quad \nonumber
\end{eqnarray}
With $0 < \theta < 1$,
\begin{eqnarray}
 P \left( Q \left(  \frac{\left \|  \bm{H} (\bm{\widehat{x}} - \bm{x}) \right \|_{2} }{\sqrt{2 N_{0}} } \right)  \geq  Q \left( \frac{ \sqrt{\theta} \left \|  \bm{\widehat{x}} - \bm{x} \right \|_{2}}{\sqrt{2 N_{0}}} \right) \right) \leq \left(\theta e^{1-\theta}\right)^{n}. \quad \nonumber
\end{eqnarray}

\emph{Proof:} Corollary 4 is established because $Q(t)$ is a monotonically decreasing function of $t$.

%
%
%
%
%
%
%

\section{Conclusion}


The asymptotic analysis of random lattices in high dimensions is presented to clarify the distance properties of lattice points. These properties indicate the asymptotic value for the distance between any pair of lattice points in high-dimension random lattices, and describe the convergence behavior of how the asymptotic value approaches the exact distance. The asymptotic analysis further prompts to new insights into the asymptotic behavior of sphere-decoding complexity and the pairwise error probability (PEP) with ML detector for a large number of antennas.




\section*{Appendix: Proofs}

\subsection{Preliminaries}
\emph{Lemma A.1}: A Gaussian random matrix $\bm{G}$ is \emph{bi-unitarily invariant}, that is, the joint distribution of its entries equals that of $\bm{U} \bm{G} \bm{V}^{\dag}$ for any unitary matrices $\bm{U}$ and $\bm{V}$ independent of $\bm{G}$, denoted by $\bm{U} \bm{G} \bm{V}^{\dag} \overset{d}{=} \bm{G}$.

\emph{Lemma A.2}: A central Wishart matrix $\bm{W}$ is \emph{unitarily invariant}, i.e., the joint distribution of its entries equals that of $\bm{V} \bm{W} \bm{V}^{\dag}$ for any unitary matrix $\bm{V}$ independent of $\bm{W}$, denoted by $\bm{V} \bm{W} \bm{V}^{\dag} \overset{d}{=} \bm{W}$.

\emph{Lemma A.3}: For two matrices $\bm{A}$ and $\bm{B}$, $\textmd{Tr}(\bm{A}^{\dag} \bm{B}) = \textmd{Vec}(\bm{A} )^{\dag} \textmd{Vec}(\bm{B} )$, where $\textmd{Tr}(\bm{A}^{\dag} \bm{B})$ is also known as the Hilbert-Schmidt inner product of $\bm{A}$ and $\bm{B}$.

If $z_{1}$, $\cdots$, $z_{k}$ are independent \emph{real} Gaussian random variables with zero mean and unit variance, then the sum of their squares, $s = \sum_{i=1}^{k} z_{k}^{2}$, is distributed according to the chi-squared distribution with $k$ degrees of freedom. This is usually denoted as $s \sim \chi_{k}^{2}$. Its cumulative distribution function (c.d.f.) is $F(x;\chi_{k}^{2}) = \frac{\gamma(k/2,x/2)}{\Gamma(k/2)}$, where $\gamma(p,x) = \int_{0}^{x} t^{p-1} e^{-t} dt$ is the lower incomplete gamma function and $\Gamma(p)$ is the gamma function (see \cite[(2.1-111)]{Proakis_Digital_1995}).

\emph{Lemma A.4}: Letting $\theta = x/k$, Chernoff bounds on the lower and upper tails of the c.d.f. of $s$ can be obtained \cite{Dasgupta:Johnson}:
\begin{itemize}
  \item For the cases when $0 < \theta < 1$ (which include all of the cases when this c.d.f. is less than half),
$F(\theta k;\chi_{k}^{2}) \leq \left(\theta e^{1-\theta}\right)^{k/2}$.
  \item The tail bound for the cases when $\theta > 1$, similarly, is $1- F(\theta k;\chi_{k}^{2}) \leq \left(\theta e^{1-\theta}\right)^{k/2}$.
\end{itemize}

\subsection{Proof of Theorem 1}

We begin the proof by taking a closer look at (\ref{pre2_thm1}).

First, we shall have $ \left \| \widetilde{\bm{G}} \bm{\Sigma}_{\Delta \bm{x}} \right \|_{F}^{2} = \textmd{Tr} \left(  \bm{\Sigma}_{\Delta \bm{x}}^{\dag} \widetilde{\bm{G}}^{\dag} \widetilde{\bm{G}} \bm{\Sigma}_{\Delta \bm{x}} \right) = \textmd{Tr} \left(  \widetilde{\bm{G}}^{\dag} \widetilde{\bm{G}} \bm{\Sigma}_{\Delta \bm{x}}  \bm{\Sigma}_{\Delta \bm{x}}^{\dag} \right) $. Then, by assuming that $\widetilde{\bm{G}} = \left[\widetilde{g}_{i,j}\right] \in C^{n \times m} $ has columns $\widetilde{\bm{g}}_{1}, \cdots,\widetilde{\bm{g}}_{m}$ and defining
$\left[\widetilde{\bm{G} }\widetilde{\bm{G}}^{\dag}\right]_{jj} \triangleq \widetilde{\bm{g}}_{j}^{\dag} \widetilde{\bm{g}}_{j}$, we can get $ \left \| \widetilde{\bm{G}} \bm{\Sigma}_{\Delta \bm{x}} \right \|_{F}^{2}  = \textmd{Vec}(\widetilde{\bm{G}}^{\dag} \widetilde{\bm{G}} )^{\dag} \textmd{Vec}\left(\bm{\Sigma}_{\Delta \bm{x}} \bm{\Sigma}_{\Delta \bm{x}}^{\dag}\right)  \quad = \left[\widetilde{\bm{G}}^{\dag} \widetilde{\bm{G}}\right]_{11} \sigma_{\Delta \bm{x}}^{2}$,
where the first equality is due to Lemma A.3. Since $ \widetilde{\bm{G}} \overset{d}{=} \bm{G}$, $\widetilde{\bm{G}}^{\dag} \widetilde{\bm{G}}$ is a central Wishart matrix and $\widetilde{\bm{G}}^{\dag} \widetilde{\bm{G}} \overset{d}{=} \bm{G}^{\dag} \bm{G}$ according to Lemma A.2. Let $\widetilde{g}_{i,j} = \widetilde{g}_{i,j,\textmd{re}} + \sqrt{-1} \cdot \widetilde{g}_{i,j,\textmd{im}}$, it follows that

\begin{eqnarray} \left[\widetilde{\bm{G}}^{\dag} \widetilde{\bm{G}}\right]_{11} = \sum_{i=1}^{n} \widetilde{g}_{i,1}^{\dag} \widetilde{g}_{i,1} = \sum_{i=1}^{n} \left( \widetilde{g}_{i,1,\textmd{re}}^{2} + \widetilde{g}_{i,1,\textmd{im}}^{2} \right). \nonumber
\end{eqnarray}

Second, by letting $\widetilde{w}_{i} = \widetilde{w}_{i,\textmd{re}} + \sqrt{-1} \cdot \widetilde{w}_{i,\textmd{im}}$,

\begin{eqnarray}
\| \widetilde{\bm{w}} \|_{F}^{2} = \sum_{i=1}^{n} \widetilde{w}_{i}^{\dag} \widetilde{w}_{i} = \sum_{i=1}^{n} \left( \widetilde{w}_{i,\textmd{re}}^{2} + \widetilde{w}_{i,\textmd{im}}^{2} \right) \nonumber
\end{eqnarray}

Third, $\textmd{Re} \left[ \widetilde{\bm{w}}^{\dag} \widetilde{\bm{G}} \bm{\Sigma}_{\Delta \bm{x}}\right] = \textmd{Re} \left[ \widetilde{\bm{w}}^{\dag} \widetilde{\bm{g}}_{1}  \sigma_{\Delta \bm{x}}\right] $, where
\begin{eqnarray}\textmd{Re} \left[ \widetilde{\bm{w}}^{\dag} \widetilde{\bm{g}}_{1} \right] = \sum_{i=1}^{n} \left( \widetilde{w}_{i,\textmd{re}}  \widetilde{g}_{i,1,\textmd{re}} +\widetilde{w}_{i,\textmd{im}} \widetilde{g}_{i,1,\textmd{im}} \right). \nonumber \end{eqnarray}

From above, (\ref{pre2_thm1}) can be rewritten as

\begin{eqnarray}
&& \left \|   \widetilde{\bm{G}} \bm{\Sigma}_{\Delta \bm{x}}  + \widetilde{\bm{w}}   \right \|_{F}^{2} = \sum_{i=1}^{n} \left[ \left( \sigma_{\Delta \bm{x}}\widetilde{g}_{i,1,\textmd{re}}+\widetilde{w}_{i,\textmd{re}} \right)^{2} \right. \nonumber \\ && \quad  + \left. \left( \sigma_{\Delta \bm{x}}\widetilde{g}_{i,1,\textmd{im}}+\widetilde{w}_{i,\textmd{im}} \right)^{2} \right]. \nonumber
\end{eqnarray}
Because the sum of two statistically independent Gaussian random variables is also a random variable, we can get $\sigma_{\Delta \bm{x}}\widetilde{g}_{i,1,\textmd{re}}+\widetilde{w}_{i,\textmd{re}} \sim \mathcal{N}\left(0, \frac{\sigma_{\Delta \bm{x}}^{2}}{2n} + \frac{N_{0}}{2}\right)$ and $\sigma_{\Delta \bm{x}}\widetilde{g}_{i,1,\textmd{im}}+\widetilde{w}_{i,\textmd{im}} \sim \mathcal{N} \left(0, \frac{\sigma_{\Delta \bm{x}}^{2}}{2n} + \frac{N_{0}}{2}\right)$. This implies $ \left( \frac{\sigma_{\Delta \bm{x}}^{2}}{2n} + \frac{N_{0}}{2} \right)^{-1} \left \|   \widetilde{\bm{G}} \bm{\Sigma}_{\Delta \bm{x}}  + \widetilde{\bm{w}}   \right \|_{F}^{2} $
is chi-squared distributed with $2n$ degrees of freedom.

By applying Lemma A.4, it follows that, with $\theta > 1$,
\begin{eqnarray} && P\left(   \left \|   \widetilde{\bm{G}} \bm{\Sigma}_{\Delta \bm{x}}  + \widetilde{\bm{w}}   \right \|_{F}^{2} \geq 2n \theta \left( \frac{\sigma_{\Delta \bm{x}}^{2}}{2n} + \frac{N_{0}}{2}  \right) \right) \nonumber \\ && \quad = 1- F(2 n \theta;\chi_{2n}^{2}) \leq \left(\theta e^{1-\theta}\right)^{n}. \nonumber \end{eqnarray}
With $0 < \theta < 1$,
\begin{eqnarray} && P\left(   \left \|   \widetilde{\bm{G}} \bm{\Sigma}_{\Delta \bm{x}}  + \widetilde{\bm{w}}   \right \|_{F}^{2} \leq 2n \theta \left( \frac{\sigma_{\Delta \bm{x}}^{2}}{2n} + \frac{N_{0}}{2}  \right) \right) \nonumber \\ && \quad = F(2 n \theta;\chi_{2n}^{2}) \leq \left(\theta e^{1-\theta}\right)^{n}. \nonumber \end{eqnarray}

Therefore, the proof of Theorem 1 is completed.

\subsection{Proof of Corollary 3}

Suppose $\theta = \theta_{LB} > 1$. By letting $A_{LB}$ denote the event $ \| \bm{G} \bm{\widehat{x}}  + \bm{w} - \bm{G} \bm{x} \|_{2}^{2} \geq  \theta_{LB} (\left \|  \bm{\widehat{x}} - \bm{x} \right \|_{2}^{2} + n N_{0} )$, we have

\begin{eqnarray}
&& E \left[N_{SC}(\bm{G},\bm{\widehat{x}},\bm{w})\right] \nonumber \\ &&  \quad = P(A_{LB}) E \left[ N_{SC}(\bm{G},\bm{\widehat{x}},\bm{w}) | A_{LB} \right] \nonumber \\ &&  \quad \quad +  P(A^{c}_{LB})
E \left[N_{SC}(\bm{G},\bm{\widehat{x}},\bm{w}) | A^{c}_{LB} \right] \nonumber \\ &&  \quad \geq P(A^{c}_{LB}) E \left[ N'(\bm{\widehat{x}},\theta_{LB}) | A^{c}_{LB} \right], \nonumber
\end{eqnarray}
where $ P(A^{c}_{LB}) = 1 - \left(\theta_{LB} e^{1-\theta_{LB}}\right)^{n}$ that tends to $1$ if $n \rightarrow + \infty$.

If $E \left[N'(\bm{\widehat{x}},\theta)\right] \overset{\smile}{\geq} e^{n \psi}$ with $\psi>0$,
\begin{eqnarray}
&& \lim_{n \rightarrow +\infty} \frac{\log  E \left[N_{SC}(\bm{G},\bm{\widehat{x}},\bm{w})\right]}{\log e^{n \psi}} \nonumber \\ && \quad \geq \lim_{n \rightarrow +\infty} \frac{\log  E \left[ N'(\bm{\widehat{x}},\theta_{LB}) | A^{c}_{LB} \right]}{\log e^{n \psi}} \geq 1, \nonumber
\end{eqnarray}
where the last equality holds due to the fact that whether the event $A^{c}_{LB}$ occurs shall not affect $N'(\bm{\widehat{x}},\theta_{LB})$. Thus, (\ref{equ_cor_3a}) can be validated.

Suppose $\theta = \theta_{UB} < 1$. Let $A_{UB}$ be the event $\| \bm{G} \bm{\widehat{x}}  + \bm{w} - \bm{G} \bm{x} \|_{2}^{2} \leq  \theta_{UB} (\left \|  \bm{\widehat{x}} - \bm{x} \right \|_{2}^{2} + n N_{0} )$. Then,

\begin{eqnarray}
&& E \left[ N_{SC}(\bm{G},\bm{\widehat{x}},\bm{w}) \right] \nonumber \\ &&  \quad = P(A_{UB}) E \left[ N_{SC}(\bm{G},\bm{\widehat{x}},\bm{w}) | A_{UB} \right] \nonumber \\ &&  \quad \quad +  P(A^{c}_{UB})
E \left[ N_{SC}(\bm{G},\bm{\widehat{x}},\bm{w}) | A^{c}_{UB} \right] \nonumber \\ &&  \quad \leq P(A_{UB}) E \left[ C_{ML} | A_{UB} \right] \nonumber \\ &&  \quad \quad +  P(A^{c}_{UB})
E \left[ N'(\bm{\widehat{x}},\theta_{UB}) | A^{c}_{UB} \right]. \nonumber
\end{eqnarray}

It is convenient to have that, as $n \rightarrow + \infty$, $P(A_{UB}) = \left(\theta_{UB} e^{1-\theta_{UB}}\right)^{n} \rightarrow 0$ and $P(A^{c}_{UB}) \rightarrow 1$. Hence,

\begin{eqnarray}
&& \lim_{n \rightarrow +\infty} \frac{\log  E \left[N_{SC}(\bm{G},\bm{\widehat{x}},\bm{w})\right]}{\log e^{n \psi_{max}}} \nonumber \\ && \quad \leq \lim_{n \rightarrow +\infty} \frac{\log \left( \left(\theta_{UB} e^{1-\theta_{UB}}\right)^{n} L^{n/\kappa} + e^{n \psi} \right)}{\log e^{n \psi_{max}}}  \leq 1, \nonumber
\end{eqnarray}
where the relation $\log(a+b) \leq  \log(2 \cdot \max\{a,b\})$ with $a,b>0$ is used for deriving the second inequality.

Therefore, (\ref{equ_cor_3b}) is proved as well.

\subsection{Proof of Theorem 2}

To prove the result, define
\begin{eqnarray}
 \mathcal{X}_{n,\alpha,L,\theta} \triangleq \quad\quad\quad\quad\quad\quad\quad\quad\quad\quad\quad\quad\quad\quad\quad\quad\quad \nonumber \\      \left \{\bm{x} \in \mathbb{CZ}^{m} \left| |\widehat{x}_{i} - x_{i}|^2 \leq \left\{ \begin{array}{c}
                                               d_{max}^2, \,\, i=1, \cdots, \eta_{n,\alpha, L,\theta} \\
                                              0, \,\, i = \eta_{n,\alpha, L,\theta}+1, \cdots,m
                                            \end{array}
  \right. \right. \right\} \nonumber
 \end{eqnarray}
where $\eta_{n,\alpha, L,\theta} = \left\lfloor \frac{(\alpha-1) n N_{0}}{\theta d_{max}^2} \right\rfloor$ and $1 < \theta < \alpha$. Since
 \begin{eqnarray}
\max_{x_{i}, \widehat{x}_{i} \in \mathcal{C}_{\tau,L}} |\widehat{x}_{i} - x_{i}|^2 = d_{max}^2, \nonumber
\end{eqnarray}
 we shall get
\begin{eqnarray}
\mathcal{X}_{n,\alpha,L,\theta}  \subset
  \left \{\bm{x} \in \mathbb{CZ}^{m} \left| \left \|  \bm{\widehat{x}} - \bm{x} \right \|_{2}^{2} + n N_{0} \leq \frac{\alpha n N_{0}}{\theta} \right. \right \}. \nonumber
 \end{eqnarray}

It is easy to have $\left |  \mathcal{X}_{n,\alpha,L,\theta} \right|  = L^{\eta_{n,\alpha, L,\theta}} \asymp  L^{\frac{(\alpha-1) n N_{0}}{\theta d_{max}^2} } $,
 and
 $E \left[N'(\bm{\widehat{x}},\theta)\right] \overset{\smile}{\geq} e^{n \cdot \frac{(\alpha-1) N_{0} \log L}{\theta d_{max}^2 }}$.

With $\theta > 1$ but arbitrarily close to $1$, from (\ref{equ_CSD_lb}) and Corollary 3, it is only a small step to the desired result:

\begin{eqnarray}C_{SD} \geq E[ N_{SC}(\bm{G},\bm{\widehat{x}},\bm{w}) ] \overset{\smile}{\geq} L^{n \cdot \min \left \{1/\kappa, \frac{(\alpha-1) N_{0}}{ d_{max}^2 } \right \} }, \nonumber\end{eqnarray}
where the last $\overset{\smile}{\geq}$ holds because $C_{SD} \leq L^{m} = L^{n/\kappa}$.

\end{document}